\documentclass[namedreferences]{solarphysics}
\usepackage[hyperref,optionalrh,solaromanenum,linksfromyear]{spr-sola-addons} 
\usepackage{solaheader}	

\usepackage{color}                       
\usepackage{breakurl}                         
\usepackage{gensymb}
\usepackage{aas_macros}
 \usepackage{booktabs}
\usepackage{graphicx}
\usepackage{subfig}
\usepackage{multirow}
\usepackage{array}

\newcommand{\solareceiveddate}{23 January 2015} 
\newcommand{\solaaccepteddate}{4 September 2015}
\begin{document}

\begin{article}

\begin{opening}

\title{An Application of the Stereoscopic Self-Similar-Expansion Model to the Determination of CME-Driven Shock Parameters}

%
\author{L.~\surname{Volpes}$^{1}$\sep
       V.~\surname{Bothmer}$^{1}$}

%
\runningauthor{L. Volpes, V.Bothmer}
\runningtitle{\footnotesize{An Application of the SSSEM to the Determination of CME-Driven Shock Parameters}}

%
  \institute{$^{1}$ Georg-August-Universit\"{a}t G\"{o}ttingen, Germany
                     \email{volpes@astro.physik.uni-goettingen.de} 
                     \email{bothmer@astro.physik.uni-goettingen.de}
             }

\begin{abstract} We present an application of the stereoscopic self-similar-expansion model (SSSEM) to \textit{Solar Terrestrial Relations Observatory} (STEREO)/\textit{Sun--Earth Connection Coronal and Heliospheric Investigation} 
(SECCHI) observations of the 03 April 2010 CME and its associated shock. The aim is to verify whether CME-driven shock parameters can be inferred from the analysis of j-maps. 
For this purpose we use the SSSEM to derive the CME and the shock kinematics. 
Arrival times and speeds, inferred assuming either propagation at constant speed or with uniform deceleration, show good agreement with \textit{Advanced Composition Explorer} (ACE) measurements.
The shock standoff distance $[\Delta]$, the density compression $[\frac{\rho_d}{\rho_u}]$ and the Mach number $[M]$ are calculated
combining the results obtained for the CME and shock kinematics with models for the shock location. Their values are extrapolated to $\textrm{L}_1$ and
compared to \textit{in-situ} data. The \textit{in-situ} standoff distance is obtained from 
ACE solar-wind measurements, and the Mach number and compression ratio are provided by the Harvard--Smithsonian Center for Astrophysics interplanetary shock database. 
They are $\frac{\rho_d}{\rho_u} =2.84$  and $M = 2.2$. The best fit to observations is obtained when the SSSEM half width $\lambda = 40 \degree$ and the CME and shock propagate with uniform deceleration.
In this case we find $\Delta = 23 \textrm{R}_{\odot}$, $\frac{\rho_d}{\rho_u} =2.61$, and $M = 2.93$.
The study shows that CME-driven shock parameters can be estimated from the analysis of time--elongation plots and can be used to predict their \textit{in-situ} values. 
\end{abstract}

%
\keywords{Coronal Mass Ejections, Interplanetary, Solar Wind, Shock Waves}

\end{opening}

%
\section{Introduction}\label{Introduction} 
Coronal mass ejections (CME) are large--scale eruptions of plasma and magnetic-field that originate from the Sun and propagate in the interplanetary medium. 
They arise above photospheric bipolar regions and travel outwards with speeds between a few hundreds and more than $2500 ~\textrm{km~s}^{-1}$
 (\citeauthor{Yashiro2004}, \citeyear{Yashiro2004}; \citeauthor{CremadesBothmer2004}, \citeyear{CremadesBothmer2004}). 
When the speed of the CME in the rest frame of the solar-wind exceeds the local Alfv\'{e}n speed, a magnetohydrodynamics (MHD) shock is driven ahead of the 
CME leading edge. Turbulence is generated in the post--shock region, the shock sheath, resulting in fluctuations of the magnetic-field and plasma parameters 
(\citeauthor{Gosling1987}, \citeyear{Gosling1987}). 
Already in the 1970s, by comparing CME speeds at a few solar radii above the Sun to the Alfv\'{e}n speed at those heights, 
it was realized that CMEs can drive shocks in the corona (see, \textit{e.g}, \citeauthor{VourlidasOntiveros2009} \citeyear{VourlidasOntiveros2009} and references therein).
With the help of today's advanced coronagraphs, under suitable circumstances, these shocks can be detected in white-light images due to the density compression ahead of the CME \citep{Vourlidas2006}. 
Although indirect evidence of the presence of shocks can be identified, for example, in metric Type--II radio bursts and in 
observations of deflected ambient streamers, it was not until 2003 that the first direct detection of a CME-driven shock was announced. 
\cite{Vourlidas2003} employed an MHD model of the observed event, and they were able to show that the bright feature in the coronagraph images
indeed corresponds to a shock in their simulation. \cite{OntiverosVourlidas2009} subsequently reported the analysis of LASCO fast events ($v > 1500 ~\textrm{km~s}{^-1}$), 
showing that in $86~ \%$ of the cases CME-driven shocks could be detected in the white-light images. Remote sensing observations of shocks can be exploited to determine shock 
parameters, such as compression ratio and Mach number \citep{Maloney2011}, and even further to estimate the coronal magnetic-field strength
(\citeauthor{GopalswamyYashiro2011}, \citeyear{GopalswamyYashiro2011}; \citeauthor{Poomvises2012}, \citeyear{Poomvises2012}). 
These estimates rely on early models for location of the Earth's bow shock derived by \cite{Spreiter1966} and  
\cite{Seiff1962}, and subsequently modified and extended to the low-Mach-number regime (\citeauthor{FarrisRussell1994}, \citeyear{FarrisRussell1994}; 
\citeauthor{RussellMulligan2002}, \citeyear{RussellMulligan2002}). \cite{GopalswamyYashiro2011} and \cite{Poomvises2012} derived the quantities necessary 
for the calculation of the shock parameters via forward modeling using respectively a circular geometry and the graduated-cylindrical-shell (GCS) model by \cite{Thernisien2006}.  \\
In the present work, we employ inverse-modeling techniques to verify whether they allow the determination of the heliospheric-shock parameters and a prediction of their \textit{in-situ} values. 
For this purpose we analyze white-light observations from the  \textit{Sun Earth Connection Coronal and Heliospheric Investigation} (SECCHI) suite onboard the NASA  
\textit{Solar Terrestrial Relations Observatory} mission (STEREO: \citeauthor{Kaiser2008},  \citeyear{Kaiser2008}; \citeauthor{Howard2008}, \citeyear{Howard2008}) of the 03 April 2010 CME and its associated shock 
(\citeauthor{Moestletal2010}, \citeyear{Moestletal2010}; \citeauthor{Liuetal2011}, \citeyear{Liuetal2011}). We separately determine the CME and shock kinematics and the time evolution of the 
standoff distance, the compression ratio, and the Mach number. Their values are then extrapolated to $\textrm{L}_1$, where they can be compared to \textit{in-situ} \textit{Advanced Composition Explorer}  
measurements (ACE: \citeauthor{Stoneetal1998}, \citeyear{Stoneetal1998}).\\
The data processing and analysis techniques employed are presented in Section 
\ref{Image_processing}. Section \ref{Methods} describes the models used in this work for the CME and shock kinematics and for the calculation of the shock parameters; 
their application to the 03 April 2010 CME is illustrated in Section \ref{Results}. In Section {\ref{discussion}} we discuss the applicability of the SSSEM for the derivation of the shock parameters.
Section \ref{conclusions} provides a summary and conclusions.

\section{Data and Image Processing}\label{Image_processing}

\begin{figure}[t] 
\centerline{\includegraphics[width=\linewidth]{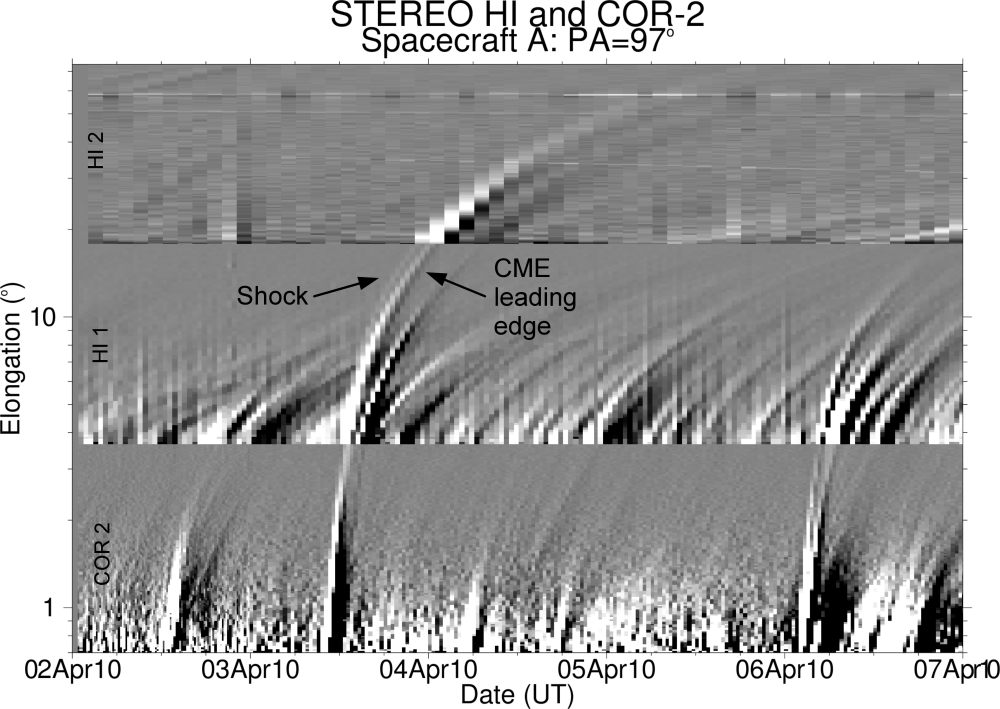}}
\caption{Time--elongation plot for the time interval between 02 April 2010 and 07 April 2010 obtained from STEREO-A COR2, HI1 and HI2
	difference images. The track corresponding to the leading edge (LE) of the CME analyzed in this work is the one appearing in the COR2 field of view on 03 April 2010 at about 09:24UT.  
        In the elongation range observed by HI1 a faint curve appears, preceding the CME. This corresponds to the density enhancement due to compression
	downstream of the shock. The tracks following the CME leading edge correspond to regions of density enhancement in the CME itself (also see Figure \ref{fig:Evolution}).
	Manual selection of points along the track corresponding to the CME and the shocked plasma allows for the independent determination
	of their kinematics.}\label{fig:Jmap}
\end{figure}

The NASA/STEREO mission, launched on 25 October 2006, consists of the twin spacecraft, 
STEREO-A (Ahead) and -B (Behind) orbiting the Sun at 1 AU. STEREO-A precedes the Earth in its orbit, while STEREO-B follows it. The separation angle 
between each spacecraft and Earth increases by about $22.5\degree$ per year. A package of five telescopes, the SECCHI suite, 
is mounted onboard both STEREO-A and -B, and provides observations of the solar corona from the solar disk to distances beyond 1 AU \citep{Howard2008}. \\
The EUVI instrument images the Sun's atmosphere in four different wavelengths, from the upper chromosphere to the corona; COR1 and COR2 are two nested 
coronagraphs with fields of view covering respectively the ranges from $1.5$ to $4 \textrm{R}_{\odot}$, and from $2.5$ to $15\textrm{R}_{\odot}$. HI1 and HI2 are heliospheric 
imagers observing the inner heliosphere up to distances larger than 1 AU. HI1 has a field of $20 \degree$, centered at $14\degree$ from the Sun 
along the Sun--Earth line; HI2 has a field of view of $70\degree$, centered at $54\degree$ along the Sun--Earth line. \\
Combined observations from the instruments of the SECCHI suite allow us
to follow the evolution of solar-wind features from their origin near the solar surface up to beyond 1 AU (\citeauthor{Davisetal2009}, \citeyear{Davisetal2009}).
Multi--point observations, moreover, provide an unprecedented insight on the three-dimensional structure and evolution of CMEs, 
and allow  the determination of their kinematics free from projection effects. 
\\
In order to use STEREO data for scientific analysis, processing of the images is required. 
Flat--field and vignetting correction, CCD bias subtraction, and correction for optical distortion are some of the calibration procedures 
applied to COR2 observations (for a complete description of COR images processing visit \href{http://hesperia.gsfc.nasa.gov/ssw/stereo/secchi/doc/cor\_prep.html}{hesperia.gsfc.nasa.gov/ssw/stereo/secchi/doc/cor\_prep.html}).
For HI1 and HI2 observations, background stray-light removal is necessary to reveal the presence of CMEs 
(\citeauthor{Daviesetal2009}, \citeyear{Daviesetal2009}; \citeauthor{Eylesetal2009}, \citeyear{Eylesetal2009}).
\\
The visibility of faint features in white-light observations can be enhanced by image differencing, a technique in which each image is obtained by subtracting two subsequent snapshots
(\citeauthor{Sheeley1999}, \citeyear{Sheeley1999}). What is shown in difference images is the change in brightness between
two consecutive snapshots: white features lead black ones in the direction of motion, the former corresponding to plasma which, in the later image, has propagated 
further away from the Sun. Time--elongation plots (also known as j-maps) are built by stacking slices of difference images cut at a fixed position angle, and they can be used to determine the
kinematics of CMEs (\citeauthor{Daviesetal2009}, \citeyear{Daviesetal2009}); structures propagating in the imagers' 
field of view appear as curved tracks in time--elongation plots (see Figure \ref{fig:Jmap}). In the interpretation of \cite{Sheeley1999} the shape of such curves is due to the apparent 
acceleration/deceleration seen by an observer measuring the angular distance (\textit{i.e.} the elongation $[\epsilon]$) of a feature that propagates away from the Sun at constant speed.

\section{Methods}\label{Methods}

\begin{figure}[t] 
\centerline{\includegraphics[width=\linewidth]{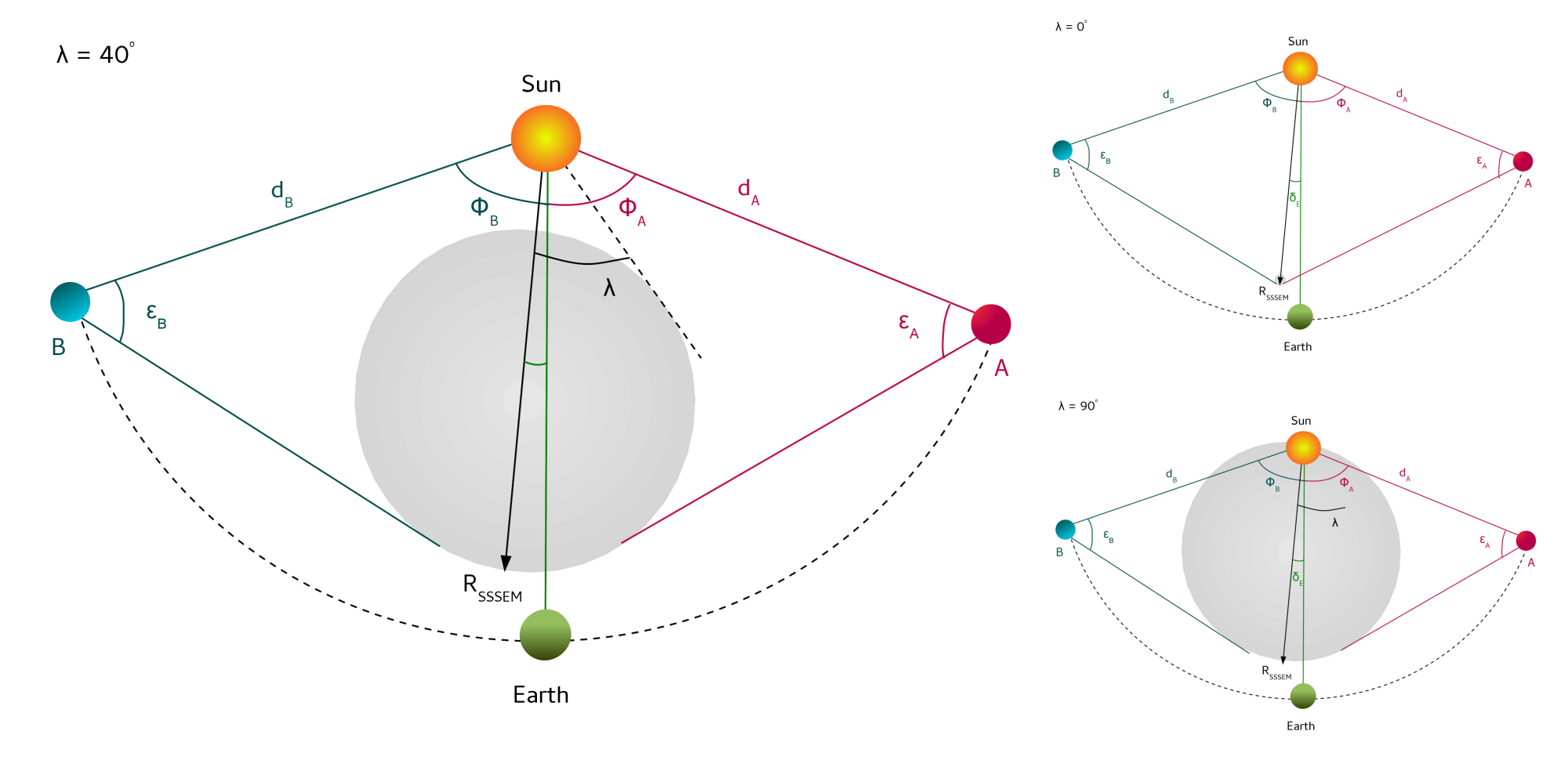}}
\caption{Schematic representation of the stereoscopic self-similar-expansion model for $\lambda =40\degree$ (left panel) and its limiting cases: the fixed-$\phi$ (top right panel)
        and the harmonic mean (bottom right panel). According to the SSSEM the CME is approximated to a sphere expanding with a constant half width $[\lambda]$. 
        The positions of the STEREO spacecraft in the pictures match the location of STEREO-A and -B at the time of observations.
        The elongations $\epsilon_{\textrm{\tiny{A}}}$ and $\epsilon_{\textrm{\tiny{B}}}$ are obtained from time--elongation plots for STEREO-A and -B observations. The Sun--spacecraft distances 
        $[d_{\textrm{\tiny{A}}}]$ and $[d_{\textrm{\tiny{B}}}]$ as well as the separation between the spacecraft, $\gamma_{\textrm{\tiny{AB}}} =\phi_{\textrm{\tiny{A}}}+\phi_{\textrm{\tiny{B}}}$ are known at all times. 
        Given the half width $[\lambda]$ as an input parameter, it is possible to determine the CME height above the Sun $[R]$ and its direction of propagation 
        with respect to STEREO-A and -B,  $[\phi_{\textrm{\tiny{A}}}$  and $\phi_{\textrm{\tiny{B}}}]$ as a function of time. The fixed-$\phi$ and 
        harmonic mean approximations are the limiting cases of the SSSEM, for $\lambda = 0\degree$ and $\lambda = 90\degree$ respectively.}\label{fig:SSEM}
\end{figure}

In order to determine the CME and shock kinematics we build j-maps from cuts of STEREO-A and -B difference images along the Ecliptic plane (see Figure \ref{fig:Jmap}). 
This allows us to determine the temporal evolution of the CME and shock height, speed, and direction of propagation in the Ecliptic plane. The results can therefore be directly compared to ACE measurements.
Manual selection of points along the curved tracks visible in the j-maps yields the evolution of the features' elongation angles as a function of time.
The time--elongation profile thus obtained can be transformed to distances from the Sun provided a set of assumptions on the CME geometry. 
In this article we use  the stereoscopic self-similar expansion model (SSSEM) developed by  \citeauthor{Daviesetal2012} (\citeyear{Daviesetal2012}, \citeyear{Daviesetal2013}), which 
 represents the CME front as a circular front expanding at a fixed angular width $[\lambda]$ (Figure \ref{fig:SSEM}). Via geometrical arguments it is possible to show that at each instant of time, 
for each spacecraft $[i]$, the distance from the Sun of the propagating feature is related to the measured elongation via

\begin{equation}\label{RSSSEM}
R = d_i\frac{\sin(\epsilon_i)(1+\sin(\lambda))}{\sin(\epsilon_i+\phi_i)+\sin(\lambda)}, \hspace{1.5cm} i=A,B,
\end{equation}

with $d_i$ being the Sun--spacecraft distance, $[\epsilon_i]$ the elongation angle obtained from the j-maps, and  $\phi_i$ being the direction of propagation of the
white-light feature with respect to the Sun--spacecraft line, while $\lambda$ is an input parameter of the model. With the additional constraint

\begin{equation}\label{gamma}
\gamma_\textrm{\tiny{AB}} = \phi_\textrm{\tiny{A}}+\phi_\textrm{\tiny{B}}
\end{equation}

where $\gamma_{AB}$ is the (known) separation angle between the two spacecraft, simultaneous STEREO-A and -B observations allow us to directly solve 
Equations (\ref{RSSSEM}) for $R$,  $\phi_\textrm{\tiny{A}}$ and $\phi_\textrm{\tiny{B}}$ and derive the kinematic properties of solar-wind features.
The direction of propagation with respect to the Sun--Earth line $[\delta_E]$ is simply derived as the difference between the Sun--spacecraft angle $[\delta_i]$ 
and the direction of propagation with respect to the Sun--spacecraft line determined from the SSSEM $[\phi_i]$.\\
The speed is obtained, using a quadratic Lagrangian interpolation, as the numerical time derivative of the radial profile \citep{Daviesetal2013}. 
It should be noted that Equation \ref{gamma}
is only valid when STEREO-A, STEREO-B the CME, Sun, and Earth lie in the same plane. This
condition is satisfied in our study since we follow the propagation of the CME and the shock in the Ecliptic plane. \\%
The model has two limiting cases: for $\lambda = 0\degree$ it coincides with the fixed-$\phi$ approximation (\citeauthor{Sheeley1999}, 
\citeyear{Sheeley1999}; \citeauthor{Rouillardetal2008}, \citeyear{Rouillardetal2008}), 
which represents the CME as a point source. As in Equation (\ref{RSSSEM}), $\phi$ is the direction of propagation with respect to
the Sun--observer line. 
The case $\lambda = 90\degree$, instead, corresponds to the harmonic mean geometry, describing the CME as a sphere anchored at the Sun \citep{Lugazetal2010}. \\
It is then possible, from time series of white-light observations, to determine the temporal evolution of outward moving features in the spacecraft field of view. 
If the CME and the shock are visible and distinguishable in white-light images and time--elongation plots, application of the SSSEM allows to separately 
derive the CME and shock kinematics. 
We applied the SSSEM for different values of $\lambda$ to the time--elongation profiles manually derived from the j-maps. We studied the interval between
$\lambda = 0\degree$ and $\lambda = 90\degree$ varying $\lambda$ in steps of $10\degree$.\\
The errors in the determined height $[\sigma_r]$ and direction of propagation $[\sigma_{\phi_{\textrm{\tiny{A}}}}]$ are obtained using the approach of \cite{Liuetal2010b}.
These are calculated via the following relations:

\begin{equation}\label{sigma_r}
\sigma_{R} = \sqrt{\left( \frac{\partial R_{\textrm{\tiny{A}}}}{\partial \epsilon_{\textrm{\tiny{A}}}}\right) \sigma_{\epsilon_{\textrm{\tiny{A}}}}^2+\left( \frac{\partial R_{\textrm{\tiny{A}}}}{\partial \epsilon_{\textrm{\tiny{A}}}}\right) \sigma_{\epsilon_{\textrm{\tiny{A}}}}^2}
\end{equation}

\begin{equation}\label{sigma_delta_E}
\sigma_{\phi_{\textrm{\tiny{A}}}} = \sqrt{\left( \frac{\partial \phi_{\textrm{\tiny{A}}}}{\partial \epsilon_{\textrm{\tiny{A}}}}\right) \sigma_{\epsilon_{\textrm{\tiny{A}}}}^2+\left( \frac{\partial \phi_{\textrm{\tiny{A}}}}{\partial \epsilon_{\textrm{\tiny{A}}}}\right) \sigma_{\epsilon_{\textrm{\tiny{A}}}}^2}.
\end{equation}

Following the approach of \cite{Liuetal2010b}, we assign an uncertainty of ten pixels to the measured elongation angles,
corresponding to $0.04\degree$ for points in COR2, $0.2\degree$ for points in HI1, and $0.7\degree$ for points in HI2. 
The derived errors are a function of the direction of propagation $[\phi_i]$, the height, the half-width $\lambda$, the positions of the spacecraft, and the instruments' resolution.
The errors in the speed and in the shock parameters are obtained by propagating the errors in $R$ and $\phi_i$.
\\
In order to infer the \textit{in-situ} arrival time and speed we fit the height--time profile to a linear 
function. The method also yields the eruption time. 
The fit is restricted to points in the HI1 field of view, where both the CME and the shock velocity can be approximated as constant. A discussion on the accuracy of constant-velocity fits to the observed elongation profiles has been given by \cite{Moestletal2014}. 
The comparison between arrival time and speed prediction based on remote-sensing observations and \textit{in-situ} measurements shows that, 
independently of the fitting method (FP, HM, SSEM), the predicted arrival speeds are higher than the observed ones, and the corresponding arrival times are earlier. 
This is supported by \cite{LugazKintner2013}, who argue that solar-wind drag, which has the effect of decelerating fast CMEs towards the solar-wind speed, 
has a substantial impact on the accuracy of arrival-time predictions. For this reason we also perform a quadratic fit to the height profile in the HI1 field
of view, to verify whether the inclusion of deceleration improves the accuracy of the \textit{in-situ} extrapolations. Arrival times and speeds are corrected accounting for the CME and shock geometry
 and direction of propagation following \cite{MoestlandDavies2013}. When the CME or the shock hit ACE, 
at a distance $d=213 \textrm{R}_{\odot}$ from the Sun, the position of their apex is determined from the angle of propagation $[\delta_E]$ and the half width $[\lambda]$ according to: 

\begin{equation}\label{Correction}
  R_{SSSEM} = d \frac{1 + \sin (\lambda)}{cos (\delta_E) + \sqrt{\sin (\lambda)^2 - \sin (\delta_E)^2}}.
\end{equation}

The importance of the correction increases with increasing $\lambda$ and $\delta_E$. We remind the reader that $\lambda \in [0 \degree, 90 \degree]$ and $\delta_E \in [-180 \degree, 180 \degree]$, 
with $\delta_E = 0 \degree$ corresponding to propagation along the Sun--Earth line.
\\
The upstream Mach number and the compression ratio can be derived from the standoff distance, \textit{i.e.} the distance between the shock front and the CME leading edge, by employing
models initially developed for the Earth's bow shock.
\cite{Seiff1962} and \cite{Spreiter1966}, based on hydrodynamic theory and experimental data, showed that for the Earth's bow shock, 
in the high-Mach-number regime, the standoff distance primarily depends on the ratio between upstream and downstream solar-wind densities $[\frac{\rho_u}{\rho_d}]$ according to

\begin{equation}\label{Spreiter_standoff}
    \frac{\Delta}{D} = 1.1\frac{\rho_u}{\rho_d}.
 \end{equation}

 $\Delta$ is the standoff distance and $D$ is the position of the nose of the magnetosphere measured from Earth, i.e the size of the dayside magnetosphere in a frame of reference with its
 origin at Earth \citep{FarrisRussell1994}.
 \\
 Using hydrodynamic relations \citep{Landau} it can be shown that the density compression ratio  $[\frac{\rho_u}{\rho_d}]$
 can be related to the upstream Mach number via
 
 \begin{equation}\label{Landau_Mach}
    \frac{\rho_u}{\rho_d} = \frac{\left(\gamma - 1\right) M^2+2}{\left(\gamma + 1\right) M^2},
   \end{equation}
with $\gamma$ being the adiabatic index. We adopted the value $\gamma = 5/3$, as in \cite{FarrisRussell1994}. \\
\cite{FarrisRussell1994} modified Spreiter's relation, replacing the size $[D]$ in Equation (\ref{Spreiter_standoff}) with the radius of curvature $[R_C]$, 

\begin{equation}\label{Farriss_standoff}
    \frac{\Delta}{R_c} = 0.81\frac{\rho_u}{\rho_d},
 \end{equation}
 arguing that the shock location would be more sensitive to the local shape of the magnetosphere, represented by the radius of curvature, rather than to its size alone.  
The change in the proportionality parameter between Equations (\ref{Spreiter_standoff}) and (\ref{Farriss_standoff}) 
results from the fact that \cite{FarrisRussell1994} estimated for the magnetosphere $\frac{R_C}{D}=1.35$. 
Furthermore, they introduced a $-1$ term in the denominator on the right hand side of Equation (\ref{Landau_Mach}),

\begin{equation}\label{Farriss_Mach}
    \frac{\Delta}{R_c} = 0.81\frac{\left(\gamma - 1\right) M^2+2}{\left(\gamma + 1\right) \left(M^2 -1\right)},
 \end{equation}
extending the applicability of Equations (\ref{Spreiter_standoff}) and (\ref{Landau_Mach}) to the low-Mach-number regime: the presence of the $-1$ term, in 
fact, allows the shock to move to infinity as the Mach number approaches unity.\\
Although a full MHD approach would be required to completely describe the CME--shock system, as well as the interaction between the solar-wind and the Earth's magnetosphere, 
it has been shown that the above approach yields good results in the high-Mach-number regime, for $M > 5$ \citep{Spreiter1966}. It is also a good approximation 
in the low-Mach-number regime, provided that $M$ is substituted for the fast magnetosonic Mach number. When the Alfv\'{e}n speed substantially exceeds the sound 
speed, moreover, the magnetosonic Mach number is not too different from the Alfv\'{e}nic Mach number $[M_{\textrm{\tiny{A}}}]$ \citep{Fairfield2001}, and $M$ can be replaced by $M_{\textrm{\tiny{A}}}$. 
Equation (\ref{Farriss_Mach}) has been used to derive CME and shock properties based on \textit{in-situ} \citep{RussellMulligan2002} and remote-sensing observations 
(\citeauthor{GopalswamyYashiro2011}, \citeyear{GopalswamyYashiro2011}; \citeauthor{Poomvises2012}, \citeyear{Poomvises2012};\citeauthor{Maloney2011}, 
\citeyear{Maloney2011}). 
The analysis of white-light images, in particular, allows the determination of shock parameters, and even of the magnetic-field strength, at distances from the
Sun where no direct plasma measurements are available.
In this work we derive the standoff distance as the difference between the shock and the CME height determined via the SSSEM.
The SSSEM, moreover, yields the CME radius of curvature as a function of the CME half-width $[\lambda]$, according to

\begin{equation}\label{RcurvSSEM}
R_C = R\frac{\sin(\lambda)}{1+\sin(\lambda)}.
\end{equation}

\begin{figure}
        \centering
        \includegraphics[width=0.3333\linewidth]{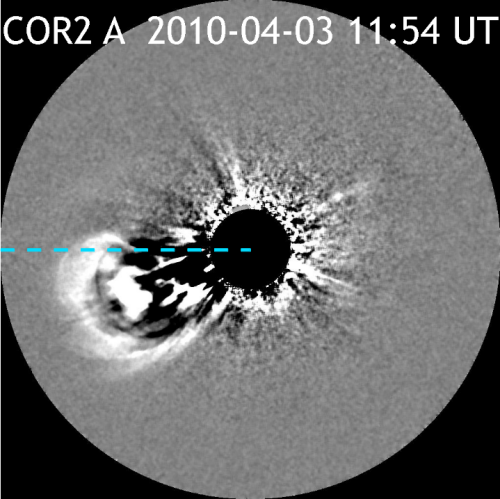}
        \includegraphics[width=0.3333\linewidth]{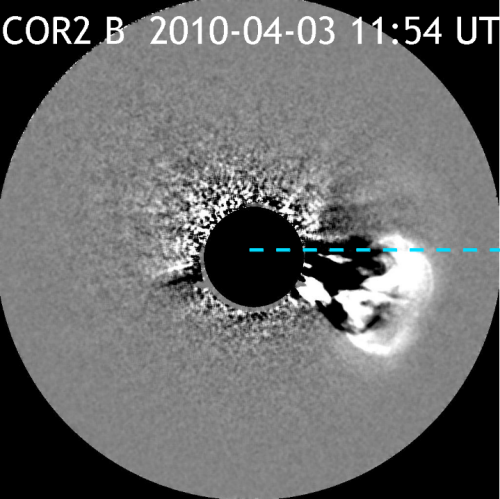}
        \vspace{1mm}
        \includegraphics[width=0.3333\linewidth]{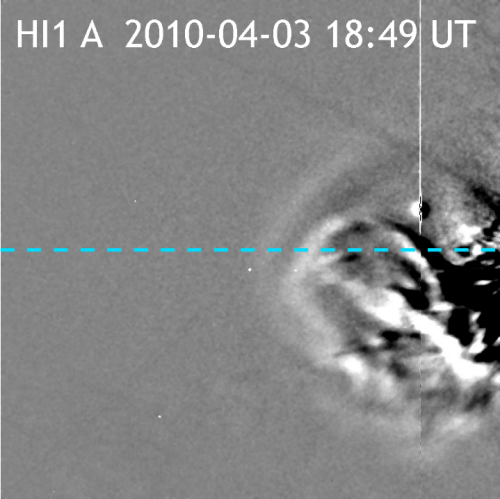}
        \includegraphics[width=0.3333\linewidth]{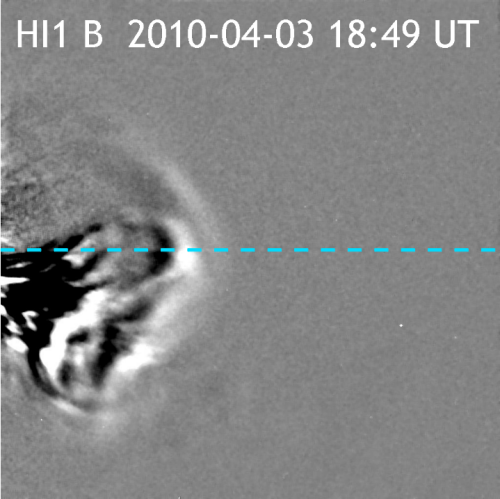}
     
	 \includegraphics[width=0.3333\linewidth]{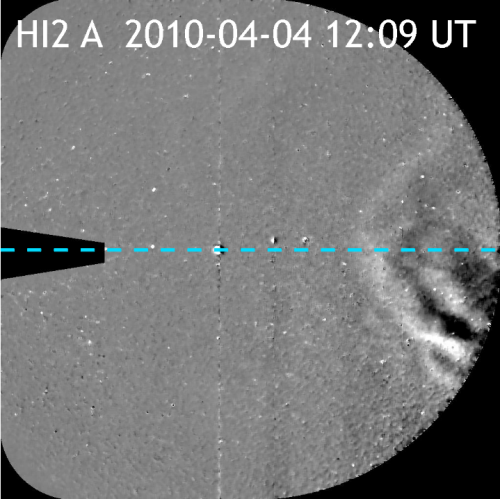}
        \includegraphics[width=0.3333\linewidth]{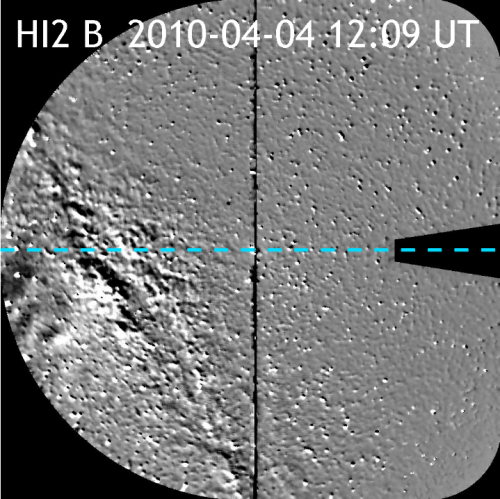}    
        \caption{a)  STEREO-A (left panel) and -B (right panel) observations of the 03 April 2010 CME. The compressed region ahead of the CME 
        leading edge is visible throughout the entire CME propagation, and particularly clear in HI1 images. 
        The Milky Way in the HI2B field of view compromises the analysis of stereoscopic observations at the elongations 
        imaged by the instrument.}
        \label{fig:Evolution}
\end{figure} 

The temporal evolution of the compression ratio and the Mach number can therefore be derived using Equations (\ref{Farriss_standoff}) and (\ref{Farriss_Mach}) via an analysis of 
j-maps.
 By assuming propagation at constant velocity or constant acceleration it is possible to extrapolate these quantities to the 
position of the ACE spacecraft, and compare them with \textit{in-situ} measurements. 
We remind the reader that, in order to determine the standoff distance at $\textrm{L}_1$, we have to take the difference between the shock and CME height corrected according to Equation (\ref{Correction}).

\begin{figure}[t]
 \centerline{\includegraphics[width=\linewidth]{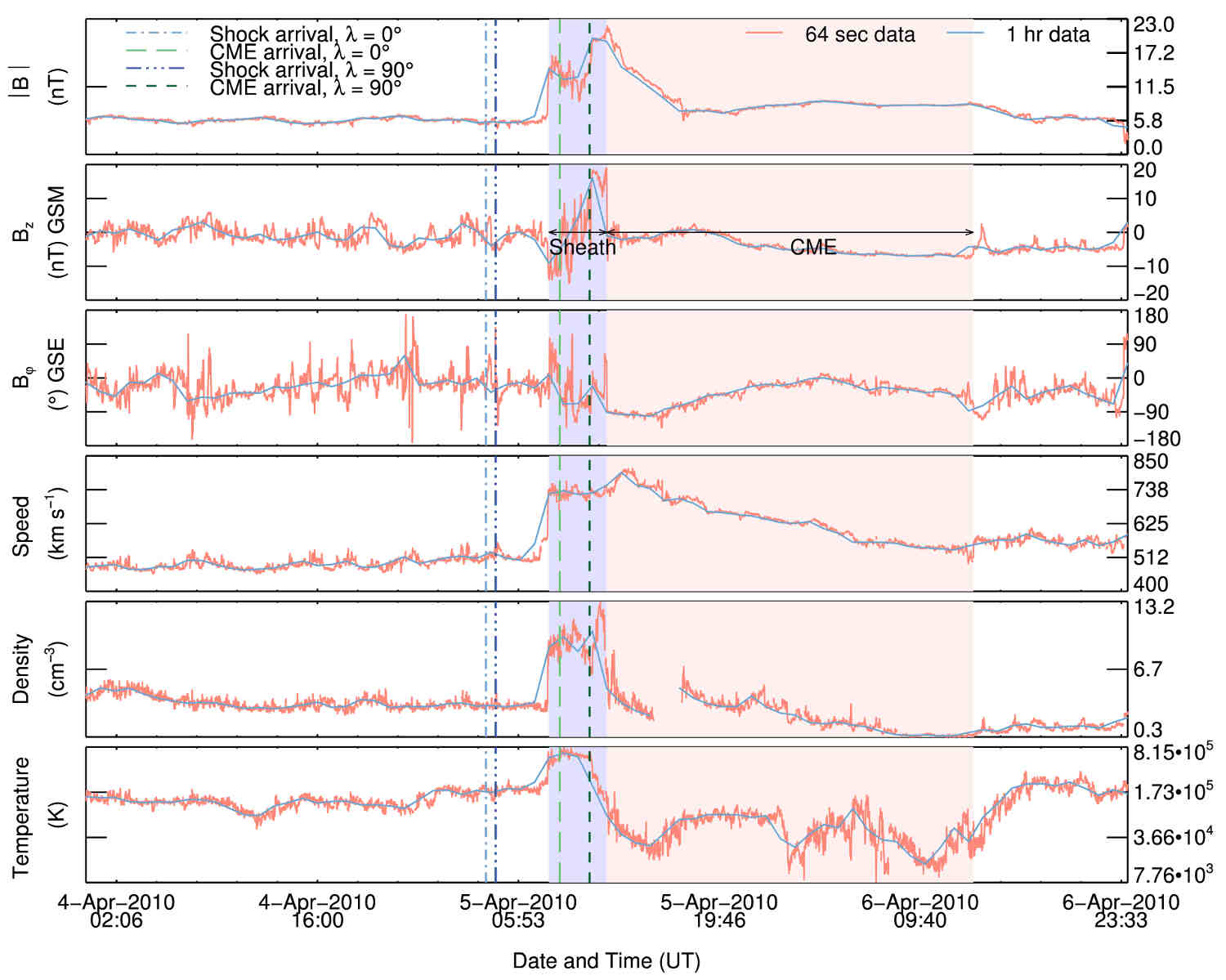}}
\caption{ACE plasma and magnetic-field measurements for the time interval 04 April 2010 to 07 April 2010. 
The shock passage is marked by the sudden increase in density and velocity at 08:00, UT on 05 April 2010; 
it was followed by the CME, which crossed the spacecraft at 12:00 and lasted for about 25 hours. 
The narrow gray area indicates the shock sheath and the wider pink area corresponds to the CME. \newline
The vertical lines represent the CME and shock-arrival times estimated via a second-degree fit to the stereoscopic self-similar 
expansion model. The light and dark dashed green lines mark the CME arrival for $\lambda =0\degree$ and $\lambda =90\degree$, respectively.
The light black dot--dashed line corresponds to the shock-arrival time for  $\lambda =0\degree$, and the dark black
dot--dashed line to the shock arrival for $\lambda =90\degree$.}\label{fig:inSitu}
\end{figure}

\section{Results}\label{Results}

The event analyzed in this work erupted from NOAA Active Region $11059$ (S$25\degree$ W$03\degree$) on 03 April 2010. 
The CME eruption was associated with a $\textrm{B7.4}$ flare detected by GOES at 09:04 UT, and resulted in a geomagnetic storm with Kp 8- on  05 April 2010 
(\citeauthor{Moestletal2010}, \citeyear{Moestletal2010}; \citeauthor{Liuetal2011}, \citeyear{Liuetal2011}). 
The separation angles between the STEREO-A and -B spacecraft and Earth at the time of observations are $67\degree$ and $71\degree$ respectively (Figure \ref{fig:SSEM}).\\
\textit{In-situ} measurements of the CME and its associated shock are provided by the ACE spacecraft: the shock passage, marked by the sudden increase in velocity 
and density, is detected on 05  April 2010 at about 08:00UT.
It is followed after about four hours by the CME. Its signatures can be identified, for instance,
in the decreasing velocity profile, in the variation of the $\phi$ component of the magnetic-field, associated to a smooth rotation of
the field itself, and in the lower temperature compared to the solar-wind (Figure \ref{fig:inSitu}). \newline
Figure \ref{fig:Evolution} shows difference images obtained from STEREO-A and -B COR2, HI1, and HI2 observations.
HI2B images are subject to high levels of background noise due to the transit of the Milky Way through the instrument's field of view. 
Stereoscopic observations of the transient are therefore only available up to 24 degrees in elongation, corresponding to the outer field of view of the HI1 imager.  
The CME morphology seems to be preserved during its outward propagation. Although the images in Figure \ref{fig:Evolution} are orthogonal to the Ecliptic plane, 
where we determine the CME and shock kinematics, the absence of significant distortion is a good indication that self-similar-expansion is a reasonable assumption for the event under study. 
This is because what we observe in white-light is the integrated election density along the line of sight.
\\
A region of enhanced density ahead of the CME, corresponding to compressed plasma downstream of the shock, can be detected in difference images of the event. 
It is especially bright in HI1 observations, and it remains visible even in the near-Earth environment imaged by HI2 A. \\
Figure \ref{fig:Jmap} shows the j-map for the event analyzed in this work. It was obtained by juxtaposing cuts of STEREO-A images along the Ecliptic plane for observations obtained
between 02 April 2010 and 07 April 2010. The elongation range in the $y$ axis covers the fields of view of COR2, which extends up to $4 \degree$, HI1 and HI2, respectively 
between about $4 \degree$ and $24 \degree$, and $19 \degree$ and $89 \degree$. 

\begin{figure}
 \centering
  {\includegraphics[width=0.48\linewidth]{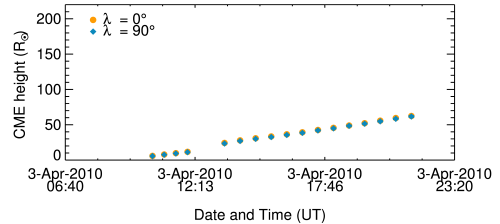}}
 {\includegraphics[width=0.48\linewidth]{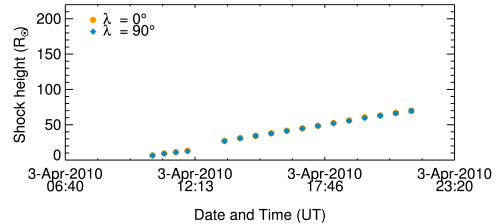}}
        \vspace{3mm}

   {\includegraphics[width=0.48\linewidth]{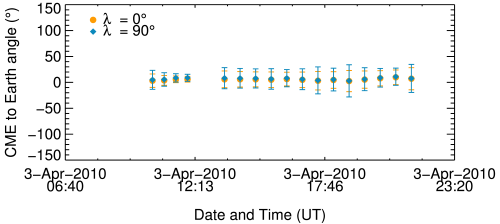}}
     {\includegraphics[width=0.48\linewidth]{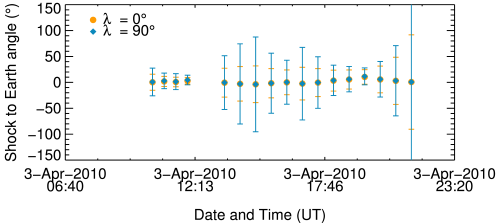}} 
        \vspace{3mm}

   {\includegraphics[width=0.48\linewidth]{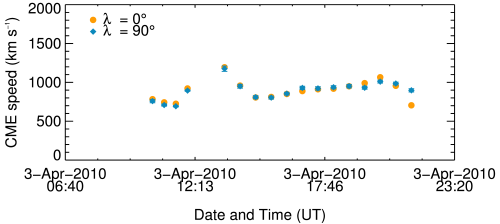}}
    {\includegraphics[width=0.48\linewidth]{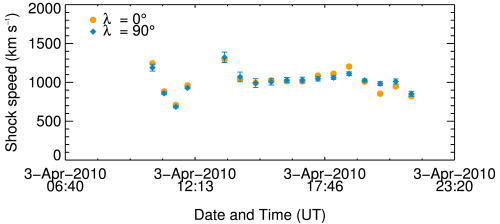}}  
            \vspace{3mm}
 
\caption{Temporal evolution of the height, direction of propagation, and velocity for the 03 April 2010 CME (left panel) and the shock (right panel), derived from 
 an application of the stereoscopic self-similar-expansion model to STEREO-A and -B COR2 and HI1 observations. The orange dots indicate solutions of the fixed-$\phi$ approximation ($\lambda =0\degree$); the 
 black dots represent the solution to the harmonic mean approximation ($\lambda =90\degree$). 
 The CME and the shock appear to propagate radially in the Ecliptic plane from the CME's source region, located at $\textrm{S}25\degree \textrm{W}03\degree$ on the Sun. }		 
\label{fig:ShockCMEkinematics}
 \end{figure}
 
\begin{figure}
 {\centering
  {\includegraphics[width=0.48\linewidth]{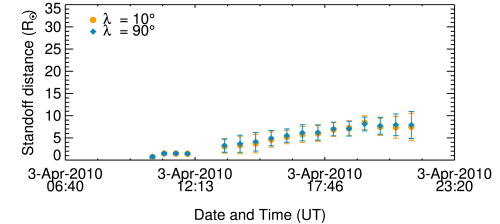}}
   {\includegraphics[width=0.48\linewidth]{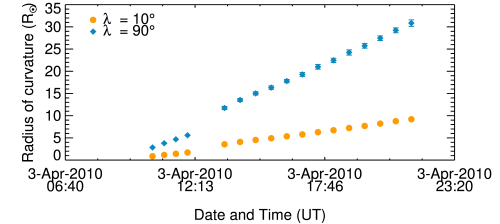}}
   {\vspace{3mm}}
  {\includegraphics[width=0.48\linewidth]{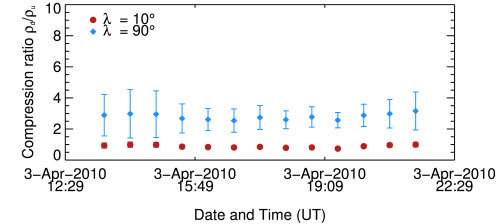}}
   {\includegraphics[width=0.48\linewidth]{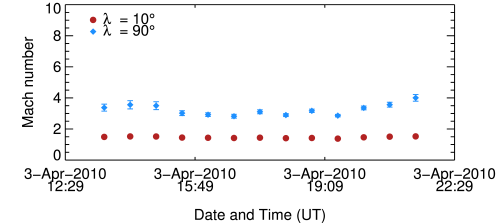}}}
   
\caption{Upper panel: standoff distance and CME radius of curvature as a function of time obtained via the application of the stereoscopic self-similar--expansion model (SSSEM) to STEREO-A and -B
COR2 and HI1 observations, for $\lambda =10\degree$ (orange dot), and $\lambda =90\degree$ (light black diamonds). 
Lower panel: temporal evolution of the compression ratio and Mach number derived according 
to Equations (\ref{Farriss_standoff}) and (\ref{Farriss_Mach}), using the values of the standoff distance and the radius of curvature obtained from
the SSSEM. The red dots represent SSSEM solutions for $\lambda =10\degree$ and the black diamonds for $\lambda =90\degree$. 
The case $\lambda = 0\degree$ is not presented, as the zero radius of curvature would lead to a singularity in the Equation (\ref{Farriss_standoff}).\newline
}
\label{fig:StandoffMach}
\end{figure}

The bright, extended curve appearing on 03 April 2010 at about 09:24 UT follows the evolution of the CME leading edge. A second, fainter curve preceding the 
CME can be identified in elongation range observed by the HI instrument. It corresponds to the propagation of the faint feature ahead of the CME
visible in HI1 images (see Figure \ref{fig:Evolution}), which we associate with compressed plasma downstream of the shock.
Since the two tracks are distinguishable from one another in the j-maps obtained from both STEREO-A and -B observations (the latter is not shown), the CME and shock kinematics could be separately determined.
This was done by applying the SSSEM to the time--elongation profile derived in the range corresponding to the COR2 and HI1 fields of view.\\
Figure \ref{fig:ShockCMEkinematics} illustrates the CME and shock height, speed, and direction of propagation with respect to Earth for the cases $\lambda =0\degree$ 
and $\lambda =90\degree$. Solutions for the intermediate values of $\lambda$ lie between the two sets of points in each plot, and are therefore not shown. 
The average direction of propagation of the CME with respect to the Sun--Earth line, $\delta_E = 5\degree \pm 2\degree$ for  $\lambda =0\degree$ and  $\delta_E = 6\degree \pm 2\degree$  for 
$\lambda =90\degree$, is compatible with radial evolution (in the Ecliptic plane) from its source region, located at W$03\degree$ latitude on the solar disk. 
These results do not differ much from the findings of \cite{Moestletal2010}, who determined $\delta_E = 9 \degree$ for $\lambda =0\degree$ and 
$\delta_E = -5 \degree$ for $\lambda =90\degree$, and $\delta_E = 0\degree \pm 2\degree$ from geometrical forward modeling. \\
Eruption time and arrival time and speed at the location of ACE are inferred by applying a first- and a second-degree fit to the CME and shock heights in the range imaged by HI1 only.
Table \ref{tab:ShockCMEkinematics2} shows a comparison between fits and observations. 
Eruption times show an excellent agreement with the time of the GOES flare detection and the time of the CME first appearance in the COR1 field of view, at 09:15 UT for both the linear and the quadratic fit. 
The overestimation of the propagation speed for both the CME and the shock in the constant-velocity fit leads to estimated arrival times earlier than the observed ACE crossing. 
Including deceleration significantly improves the arrival-time and speed predictions, particularly for the $\lambda = 90 \degree$ case. 
Based on the CME and shock heights obtained from the SSSEM results, the standoff distance, the compression ratio, and the Mach number can be computed according 
to Equations (\ref{Farriss_standoff}), (\ref{Farriss_Mach}), and (\ref{RcurvSSEM}), substituting $\gamma = 5/3$. Their time profiles and the evolution of 
the standoff distance and the radius of curvature are shown in Figure \ref{fig:StandoffMach} for  $\lambda = 0\degree$ and  $\lambda = 90\degree$. 

\begin{table}[t]
\tiny
\begin{tabular}{lllll}
\hline
                    & ACE+EUVI                   & 1st degree fit $\lambda = 0\degree$          & 1st degree fit $\lambda = 90\degree$ \\ \hline
CME eruption time   &  {03 Apr. 2010 09:04} &  {03 Apr. 2010 08:24}                   &  {03 Apr. 2010 08:32}           \\
Shock arrival time  &  {05 Apr. 2010 07:56} &  {05 Apr. 2010 00:26}                   &  {05 Apr. 2010 00:19}           \\ 
Shock arrival speed &  {750 $km~s^{-1}$}   &   {1025 $km~s^{-1}$}                &  {1033 $km~s^{-1}$}                  \\ 
CME arrival time    &  {03 Apr. 2010 12:24} &  {05 Apr. 2010 05:24}                   &  {05 Apr. 2010 06:05}           \\ \hline \\ \hline

                    & ACE+EUVI                   & 2nd degree fit $\lambda = 0\degree$      & 2nd degree fit $\lambda = 90\degree$ \\ \hline
CME eruption time   &  {03 Apr. 2010 09:04} &  {03 Apr. 2010 08:27}               &  {03 Apr. 2010 08:12}           \\
Shock arrival time  &  {05 Apr. 2010 07:56} &  {05 Apr. 2010 03:38}               &  {05 Apr. 2010 04:19}           \\ 
Shock arrival speed &  {750 $km~s^{-1}$}         &  {867 $km~s^{-1}$}                       &  {851 $km~s^{-1}$}                   \\ 
CME arrival time    &  {03 Apr. 2010 12:24} &  {05 Apr. 2010 08:44}               &  {05 Apr. 2010 10:48}           \\ \hline

\end{tabular}
\caption{Upper panel: CME arrival and eruption time, and shock-arrival time and speed determined via a constant-speed fit to the 
self-similar-expansion model (SSSEM) results for the HI1 fields of view. 
ACE \textit{in-situ} measurements of the shock and CME arrival are reported for comparison, as well as the time of flaring activity. In both the cases $\lambda =0\degree$ (FP approximation) and $\lambda =90\degree$ (HM approximation) the derived eruption times are very close
to the observed flare time, with a difference of about 30 minutes for the FP results and 1 hour for the HM results. Predicted arrival times are roughly eight hours earlier 
than the time of CME and shock passage at the location of ACE, and consistently the speeds are higher. Lower panel: Same quantities as above, obtained 
via a constant-negative-acceleration fit to the SSSEM results. The eruption times are again in excellent
agreement with observations, and a significant improvement of the arrival-time and speed prediction is found, with accuracy as high as two hours. The inferred speeds
are also closer to the ones determined through the linear fit.}
\label{tab:ShockCMEkinematics2}
\end{table}

The case  $\lambda =0\degree$ is excluded, as the corresponding zero curvature leads to a singularity in Equation (\ref{Farriss_standoff}). 
It should be pointed out that the compression ratio $[\frac{\rho_d}{\rho_u}]$ plotted in Figure (\ref{fig:StandoffMach}) is the inverse 
of the ratio on the left-hand side of Equations (\ref{Spreiter_standoff}) and (\ref{Farriss_standoff}). Both the density compression and the Mach number 
appear to be roughly constant during the CME evolution, in agreement with the results found by \cite{Maloney2011}. The increase of the standoff distance 
is also in good agreement with observations, which show the shock progressively moving away from its driver. 

An extrapolation of the standoff distance, the compression ratio, and the Mach number to the location of ACE was performed assuming again propagation at constant speed as well as at constant acceleration.  
The correction in Equation (\ref{Correction}) was applied to the CME and shock heights to determine the standoff distance at $\textrm{L}_1$.
For the $\lambda =10\degree$ case we find an unphysical compression $\frac{\rho_d}{\rho_u}  <1$; this can be caused by the overestimated value of the standoff distance, or the 
small CME curvature associated to this value of the half width.
Validation is necessary for the extrapolated values of compression ratio and the Mach number at 1 AU. The \textit{in-situ} standoff distance is calculated directly
from the CME and shock-arrival time and speed as measured by ACE. The extrapolated values of the density compression ratio and the Mach number are instead
compared to the results published on the interplanetary (IP) shock database of the Harvard-Smithsonian Center for Astrophysics (\href{https://www.cfa.harvard.edu/shocks/}{www.cfa.harvard.edu/shocks/}).
The shock parameters are determined applying different analysis methods and principles such as, \textit{e.g}, the magnetic and velocity coplanarity theorem and the solution of the 
Rankine--Hugoniot equations, to the ACE and Wind plasma and magnetic-field measurements. We use the average of the results
obtained with the different methods.\\
Table \ref{tab:MachStandoff1} shows the values of the standoff distance, Mach number, and density compression extrapolated to 1 AU
via linear and quadratic fit to the SSSEM results, as well as those calculated from \textit{in-situ} data. 
The \textit{in-situ} standoff distance is $\Delta_{\textrm{\tiny{ACE}}} = 19 \textrm{R}_{\odot}$. A comparison with the linear and quadratic fits to the SSSEM results, 
shows that for all the values of $\lambda$ the extrapolated standoff distance, ranging from $21 \textrm{R}_{\odot}$ to $30 \textrm{R}_{\odot}$, overestimates 
the observed one. This is easily recognizable in Figure \ref{fig:inSitu}, where the measured and the extrapolated magnetosheath thicknesses can be directly compared. 
\\
The \textit{in-situ} value of the Mach number, $M_{\textrm{\tiny{ACE}}} = 2.2$ falls within the ranges determined via the linear and quadratic fit, $1.38<M< 3.50$ and $1.40<M< 3.54$, 
respectively. The same is true for the compression ratio, whose \textit{in-situ} value $\frac{\rho_d}{\rho_u}_{\textrm{\tiny{ACE}}} = 2.84$ is contained in the intervals 
derived from first and second-degree fits, $0.75 <\frac{\rho_d}{\rho_u} < 2.97$,  $0.78 <\frac{\rho_d}{\rho_u} < 2.91$.
The value of $\lambda$ that best approximates the \textit{in-situ} Mach number and compression is $\lambda = 40\degree$ for both the linear and quadratic fit to the SSSEM height. For the 
 linear fit we find $\Delta = 27 \textrm{R}_{\odot}$, $\frac{\rho_d}{\rho_u} = 2.19$ and  $M = 2.42$, while for the quadratic fit we find $\Delta = 23 \textrm{R}_{\odot}$, $\frac{\rho_d}{\rho_u} = 2.61$ and  $M = 2.93$. 
 As for the arrival time we find a significant improvement in the \textit{in-situ} extrapolations when deceleration is included in our fits.

\begin{table}[t]
\tiny
\begin{tabular}{lllll}
\hline
                    & ACE                         & 1st  degree fit $\lambda = 10\degree$            & 1st  degree fit $\lambda = 90\degree$       \\ \hline
Standoff distance    &   {19 $\textrm{R}_\odot$}  &   {$30$   $\textrm{R}_\odot$}                    &   {$26$  $\textrm{R}_\odot$ }                \\
Mach number            &   $2.2$                &   {$1.38$}                                 &   {$3.26$}                             \\
Compression ratio     &   $2.84$                &   {$0.75$}                                 &  {$2.82$}                    \\    \hline      \\  \hline

                  & ACE                         & 2nd  degree fit $\lambda = 10\degree$            & 2nd  degree fit $\lambda = 90\degree$       \\ \hline
Standoff distance    &   {$19$ $\textrm{R}_\odot$}  &   {$29$  $\textrm{R}_\odot$}                    &   {$26$  $\textrm{R}_\odot$ }                \\
Mach number               &   {$2.2$}      &   {$1.41$}                                 &   {$3.46$}                             \\
Compression ratio     &   {$2.84$}         &   {$0.78$}                                 &  {$2.91$}                              \\ \hline 

\end{tabular}
\caption{Upper panel: extrapolation to 1 AU of the standoff distance, the Mach number, and the compression ratio.  
The results have been obtained via first-degree fit to the CME and shock height--time profile determined from the self-similar--expansion model
(SSSEM) applied to STEREO-A and -B HI images,
for $\lambda =10\degree$ and $\lambda = 90\degree$. 
The value of the standoff distance determined from \textit{in-situ} ACE measurements and those of the Mach number and compression ratio available on
the Harvard-Smithsonian Center for Astrophysics interplanetary shock database are reported for comparison.
Lower panel: same quantities as above, determined via a second-degree fit to the CME and shock height.
The standoff distance derived via the SSSEM is higher than the observed one. The values of the compression ratio and the Mach number computed from \textit{in-situ} measurements are within the range defined by our results. 
For $\lambda =10\degree$, the unphysical 
result $\frac{\rho_d}{\rho_u}<1$ can be related to the high standoff distance or to the small associated CME curvature.}
\label{tab:MachStandoff1}
\end{table}

\begin{figure}[t]	
  \centering
    \subfloat{%
    \includegraphics[width = 0.25\linewidth]{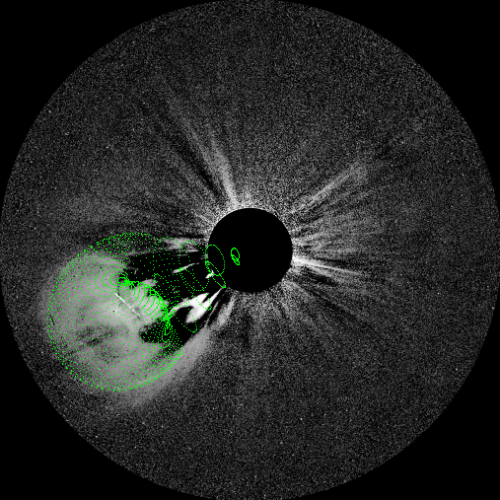}
      }
      \subfloat{%
    \includegraphics[width = 0.25\linewidth]{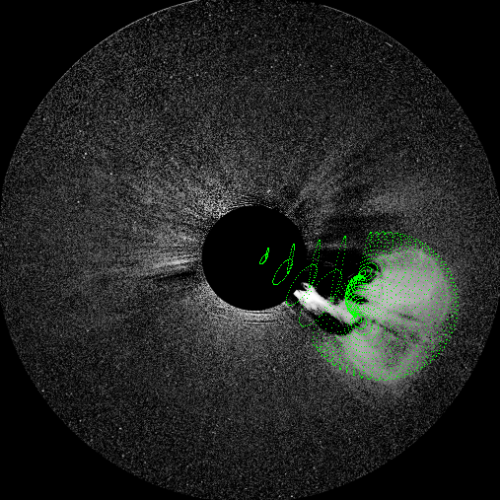}
      }\\
      \subfloat{%
    \includegraphics[width = 0.25\linewidth]{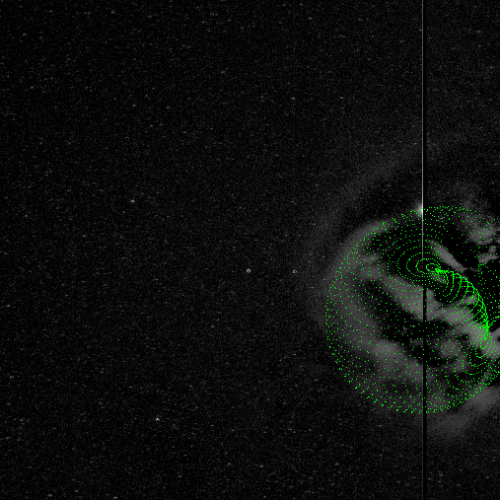}
      }\subfloat{%
    \includegraphics[width = 0.25\linewidth]{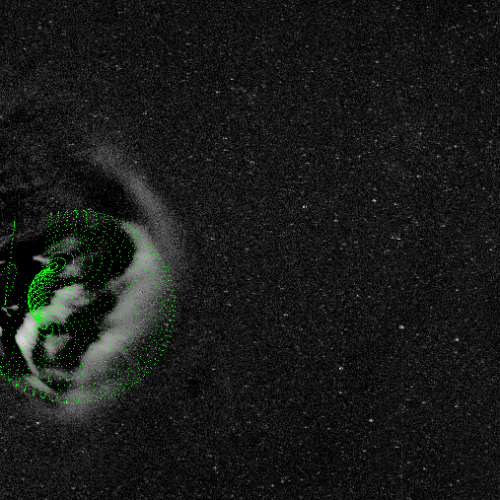}
      }
   \caption{Graduated-cylindrical-shell fitting for the 03 April 2010 CME for the COR2 and HI1 fields of view. The fit performed in COR2 was extended to HI1 assuming self-similar-expansion. The CME propagates at constant longitude 
   and latitude. The latter was used as input for the assessment of the errors associated to the calculation of the shock parameters from the standoff distance computed via the SSSEM. }
    \label{GCS}
  \end{figure}
 
\section{On the Errors Associated to Off-Ecliptic Propagation}\label{discussion}

\begin{figure}[t]	
  \centering
    \subfloat[\label{SSSEM_corrections}]{%
    \includegraphics[width = 0.32\linewidth]{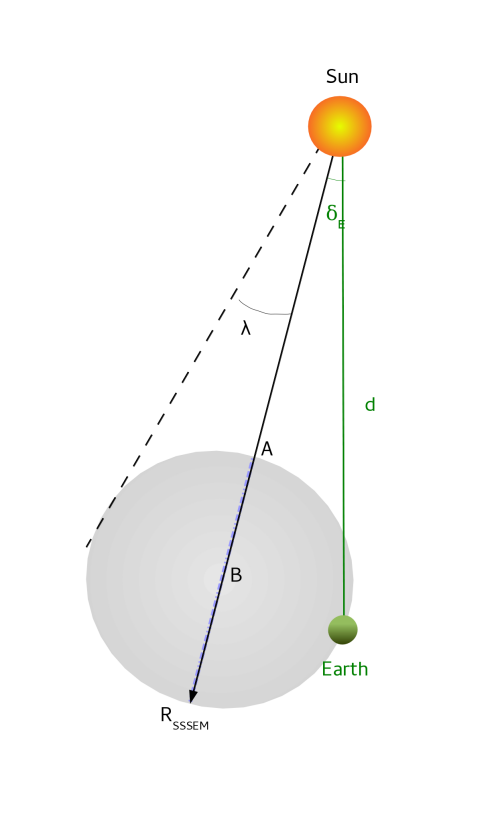}
      }
      \subfloat[\label{SSSEM_corrections2}]{%
    \includegraphics[width = 0.65\linewidth]{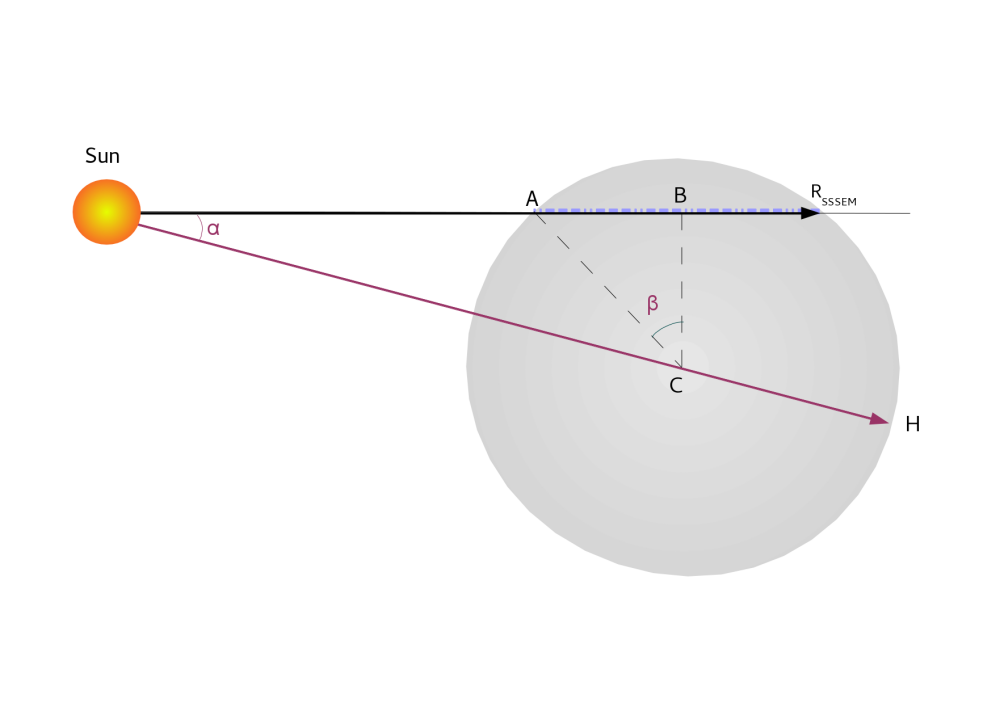}
      }
      \hfill
   \caption{Left panel: CME propagation along the Ecliptic plane, as in Figure \ref{fig:SSEM}. Right panel: CME propagation perpendicular to the Ecliptic plane, with the horizontal axis aligned to the CME height vector. The 
   plane intersects the CME, assumed to be a sphere, at its poles.}
    \label{SSSEM_corrections_all}
  \end{figure}

In order to derive the Mach number and the compression ratio using the relations in Equations (\ref{Farriss_standoff}) and (\ref{Farriss_Mach}), the standoff distance
has to be measured at the CME nose. Therefore attention should be paid to the direction of propagation of the CME when employing the SSSEM for the determination of shock parameters. 
If the direction of propagation of the CME with respect to the Ecliptic plane $[\alpha]$ is not small the determination of shock parameters using the SSSEM might lead to incorrect results. The standoff 
distance would be measured at the wrong location and the SSSEM radius of curvature might not well represent the one at the nose of the CME. Furthermore, in order to compare the SSSEM results to \textit{in-situ} data, 
we computed the standoff distance from the SSSEM heights corrected according to Equation (\ref{Correction}). In this way we determined the shock parameters from 
the standoff distance calculated along the Sun--Earth line. It is now necessary to estimate how much these values deviate from those computed at the CME nose. \\
From STEREO observations (see Figure \ref{fig:Evolution}) it is evident that the CME propagates southwards both in COR2 and in HI1. To estimate its direction of propagation in 3D we model the CME 
and the shock with the graduated-cylindrical-shell (GCS) model by \cite{Thernisien2006}. The model, which describes the CME as a croissant-shaped flux rope expanding self-similarly, 
has previously been employed to determine the shock parameters  (\cite{Poomvises2012}). We start from results obtained in COR2, and extend them to the HI1 fields of view. 
We expand the GCS flux rope keeping the aspect ratio and the half angle constant, \textit{i.e.} assuming self-similar propagation. We find that the CME propagates southwards at an angle of $\alpha = -23 \degree$ with the Ecliptic plane 
(see Figure \ref{GCS}), and at an angle of about $\delta_E^{GCS} = 5 \degree$ with respect to the Sun--Earth line. This is in agreement with our SSSEM results, which yield $\delta_E \approx 1 \degree$ for the shock  
and $\delta_E \approx 5 \degree$ for the CME. No deflection is observed during the CME propagation, \textit{i.e.} the CME longitude $\delta_E^{GCS} $ and latitude 
$[\alpha]$ are the same in the COR2 and HI fields of view. We use the value of $\alpha$ determined from the GCS model and that of $\delta_E $ obtained from the SSSEM to estimate the difference between the standoff distance 
and the CME curvature computed at the nose and 
the ones derived from the SSSEM. We assume $\alpha$ to be the same for the CME and the shock. Finally, we determine the Mach number and compression ratio at the CME nose and compare them to the results 
in Table \ref{tab:MachStandoff1}.\\
To estimate the value of the standoff distance at the CME nose we assume that the CME is a sphere of radius $R$, as this is the straightforward extension of the 
SSSEM to three-dimensional space. We assume that the intersection of the sphere with the Ecliptic plane is the SSSEM circle (see Figure \ref{SSSEM_corrections}). 
We now consider the plane formed by the CME height vector (\textit{i.e.} the vector originating at the Sun and ending at $R_{SSSEM}$) and perpendicular to the ecliptic. 
The plane cuts the sphere at the poles, so that the projection of the sphere on this is a circle of radius $R$ (see Figure \ref{SSSEM_corrections2}).
The SSSEM circle in this plane degenerates to a segment of length $2 r$, where $r$ is the radius of the SSSEM circle, \textit{i.e.}

\begin{equation}\label{r}
  r = R_{SSSEM}\frac{\sin (\lambda)}{1+\sin (\lambda)}.
\end{equation}

The radius of the sphere, which corresponds to the \textquotedblleft true\textquotedblright CME curvature, is given in terms of $r$ according to 

\begin{equation}\label{R}
   R = \frac{r}{\sin (\beta)},
\end{equation}
(see Figure \ref{SSSEM_corrections2}).\\ 
The angle $\beta$ can be determined considering that $\overline{BC} = \overline{AB} \ \tan \beta$, with $\overline{AB} = r$, and that 
$\overline{BC} = \overline{OB} \ \tan \alpha$, with $\overline{OB} = R_{SSSEM} - r$. We get

\begin{equation}\label{tanbeta}
  \tan \beta = \frac{r}{(R_{SSSEM} -r) \tan \alpha}.
\end{equation}

Finally, the CME apex is at a distance $H = \overline{OC}+R$ given by

\begin{equation}\label{H}
 H = \frac{(R_{SSSEM} -r) }{cos \alpha} + \frac{r }{\sin \beta}.
\end{equation}

At a fixed latitude, the error in neglecting the off-Ecliptic propagation will decrease with increasing $\lambda$, \textit{i.e.} with increasing size of the CME.
This is because the horizontal axis in Figure \ref{SSSEM_corrections2} will slice the sphere closer to its Equator for larger CMEs. At fixed $\lambda$, on
the other hand, the error will increase with increasing $\alpha$, as the slice will be cut closer to the poles, \textit{i.e.} further away from the apex.\\
It is possible then to derive the standoff distance and the CME curvature at the CME nose, compute the shock parameters, and compare them to our \textit{in-situ} extrapolations. 
The results are summarized in Table \ref{tab:sssemcorr} for the case $\lambda = 40 \degree$, which corresponds to our best fit to the data. 
For each parameter $X$ in the Table (\textit{i.e.} standoff distance, curvature, compression ratio, and Mach number) we report the value of 
$\frac{X_{\textrm{\tiny{corr}}} - X}{X}$, $X_{\textrm{\tiny{corr}}}$ being the parameter calculated at the CME apex.

\begin{table}[t]
\tiny
\begin{tabular}{lll}
\hline

{Relative error for:}  & {1st  degree fit $\lambda = 40\degree$}            &{ 2nd  degree fit $\lambda = 40\degree$ }      \\ \hline
{Standoff distance}    &   { {$5~ \%$}}	 			 	 &   { {$4~ \%$}}                   \\
{Curvature }           &   { {$20~ \%$} }         			 &   { {$20~ \%$}}                             \\
{Mach number}            &   { {$13~ \%$} }          			 &   { {$23~ \%$}}                             \\
{Compression ratio}    &   { {$13~ \%$} }            			 &  {  {$15~ \%$}}                             \\    \hline

\end{tabular}
\caption{{Relative difference in the values of the shock parameters calculated along the Sun--Earth line and at the CME apex for the best fit to \textit{in-situ} observations
 obtained via application of the SSSEM.}}
\label{tab:sssemcorr}
\end{table}

{The radius of curvature is the quantity that is affected the most by neglecting the off-Ecliptic propagation of the CME and the shock.
The apex value of the standoff distance is at most 5$~ \%$ bigger than the one we determined with the SSSEM. The difference between the SSSEM values of the Mach number and compression ratio computed 
at the CME nose and via the SSSEM  are between 23$~ \%$ and 13$~ \%$, with the second-degree fit being affected the most. 
This indicates that the results we obtained by computing the standoff distance via the SSSEM do not deviate dramatically from those obtained calculating the standoff distance and the curvature of the CME at its nose.}
%

\section{Summary and Conclusions}\label{conclusions}
It has been long believed that fast CMEs can drive shocks in the corona, and that their signatures could be observed in coronagraph images due to the 
density compression across the shock. 
Thanks to the increasing instrument sensitivity it is now possible to detect CME-driven shocks in a vast number of events 
\citep{OntiverosVourlidas2009}. Stereoscopic observations, moreover, provide an unprecedented insight into the 3D properties of coronal mass ejections, and 
allow the determination of CME kinematics without the need to assume constant velocity and strictly radial propagation (\citeauthor{Liuetal2010a}, \citeyear{Liuetal2010a};
\citeyear{Liuetal2010b}). In this work we present the analysis of the fast CME that erupted on 03 April 2010, 
associated with an interplanetary shock detected by ACE on 05 April 2010. The results of this work show that the shock was already present in the corona below $15 \textrm{R}_{\odot}$, 
and was visible in white-light images throughout the whole evolution towards Earth. Exploiting the fact that the CME and the shock are visible and distinguishable 
in time--elongation plots, we derived the CME and shock kinematics independently, employing the SSSEM for different values of the angular width. 
To our knowledge, this is the first time that the analysis of j-maps has been applied separately to a CME and its driven shock. 
The results of our study show that the CME evolved radially in the Ecliptic plane from the eruption site located at S$25\degree$ W$03\degree$ on the solar disk. Linear and second-degree fits to 
the height profile were performed, showing that the latter yields better estimates for arrival times and speeds. 
The standoff distance, the compression ratio, and the Mach number evolution were derived; their values were extrapolated to the location of 
ACE by exploiting a linear and a quadratic fit to the CME and shock heights. In order to validate these results,
the standoff distance was computed based on the \textit{in-situ} CME and shock arrival times and speeds. 
The values of the Mach number and the compression ratio were compared to results published on the CfA IP shock database, 
computed from ACE plasma and magnetic-field data, according to which $M = 2.2$ and $\frac{\rho_d}{\rho_u} = 2.84$. 
We found the case $\lambda = 40\degree$ to best match observations. If constant velocity is assumed, $\Delta = 27 \textrm{R}_{\odot}$, $\frac{\rho_d}{\rho_u} = 2.19$ and  $M = 2.42$. 
When deceleration is assumed to take place in the CME propagation, we find $\Delta = 23 \textrm{R}_{\odot}$,  $\frac{\rho_d}{\rho_u} = 2.61$ and  $M = 2.93$. \\
Although the CME apex was at an angle $\alpha = -23 \degree$ with respect to the Ecliptic plane, an assessment of the error associated 
to the use of the SSSEM to determine the shock parameters showed that their values at the CME nose do not substantially differ from those obtained via the SSSEM.\\
The methods presented in this work allow the determination of CME-driven shock properties from remote-sensing observations, where \textit{in-situ} measurements
are not available. Magnetic-field strength estimates are also possible based on such results \citep{GopalswamyYashiro2011}, which overcome the 
difficulty of directly measuring the coronal magnetic-field. \\
The study shows that it is possible to use inverse-modeling techniques to determine the time evolution of CME-driven shock parameters and to predict
their \textit{in-situ} values.
The NASA \textit{Solar Probe Plus} mission to be launched in 2018 will explore the near Sun environment at distances
up to $9.86 \textrm{R}_{\odot}$, \textit{i.e.} at $0.046$ AU, and the ESA Solar Orbiter mission, also to be launched in 2018, will orbit the Sun at a distance of
$0.28 AU$. The white-light observations from the \textit{Wide-Field Imager on Solar PRobe Plus} (WISPR: \citeauthor{Vourlidas2015}, \citeyear{Vourlidas2015}) and  \textit{Solar Orbiter Heliospheric Imager} 
(SoloHI: \citeauthor{Howard2013}, \citeyear{Howard2013})
cameras coupled with \textit{in-situ} measurements will provide unprecedented knowledge about the coronal plasma and magnetic-field properties at locations
unexplored so far.
Application of the models employed in this work to CME images acquired  by WISPR and SoloHI will give 
insight on the spatial variation of CME-driven shock properties. At the same time, a deeper knowledge of the plasma and magnetic-field characteristics at the same locations will help to further validate the model results.

%
\begin{acks}
The authors acknowledge funding by Deutsche Forschungsgemeinschaft (DFG) under grant SFB 963/1, project A2. They are further grateful for support in the analysis of STEREO SECCHI data 
by Jackie Davies and Richard Harrison from Rutherford Appleton Laboratory, Didcot, Chilton, UK, for the information on shock databases provided by Luciano Rodriguez from the Royal Observatory
of Belgium, Bruxelles, and for the GCS modeling in COR2 by Eckhard Bosman from the University of G\"ottingen. Volker Bothmer acknowledges support of the CGAUSS (Coronagraphic German And US Solar Probe Plus Survey) 
project for WISPR by the German Space Agency DLR under grant 50 OL 1201.

\end{acks}


%
%

\end{article} 
\end{document}